\documentclass[reprint,amsmath,amssymb,aps,superscriptaddress]{revtex4-2}

\usepackage{bbm}
\usepackage{graphicx}
\usepackage{dcolumn}
\usepackage{bm}
\usepackage[normalem]{ulem}

\usepackage{booktabs}
\usepackage{tabularx}

\usepackage{comment}

\usepackage{float}
\usepackage{todonotes}

\usepackage[colorlinks, linkcolor = black, citecolor = black, filecolor = black, urlcolor = black]{hyperref} 
\usepackage{hyperref}

\begin{document}


\title{
Electrical conductivity of the quark-gluon plasma from the low energy limit of photon and dilepton spectra}

\author{Stefan Floerchinger}
\affiliation{Institut für Theoretische Physik, Universität Heidelberg, 69120 Heidelberg, Germany}
\author{Charlotte Gebhardt}
\affiliation{Institut für Theoretische Physik, Universität Heidelberg, 69120 Heidelberg, Germany}
\author{Klaus Reygers}
\affiliation{Physikalisches Institut, Universität Heidelberg,
Im Neuenheimer Feld 226, 69120 Heidelberg, Germany}


\begin{abstract}
Fluid dynamic considerations are used to determine the electric current spectral density in the regime of small energies and momenta. The spectral density in this regime is parameterized by the electric conductivity, the charge susceptibility, and the relaxation time for the electric current, which is needed for relativistic causality. Experimentally, the spectral function can be accessed through the production rates of photons and dileptons in the expanding quark-gluon plasma. We use fluid dynamic simulations of high energy nuclear collisions, together with the transport limit of the spectral density, to obtain photon and dielectron spectra for different values of the conductivity and relaxation times. The yields of photon and dileptons produced in the plasma are compared to the background from decays of short-lived hadrons. We discuss how experiments can constrain the electrical conductivity and associated relaxation time of the quark-gluon plasma.
\end{abstract}

\maketitle

\section{Introduction}
The quark-gluon plasma (QGP) is a highly interesting state of matter that has filled the universe shortly after the Big Bang and that can be investigated in the laboratory using high energy nuclear collisions. Thermal equilibrium and close-to-equilibrium transport properties are fully determined by the microscopic properties of QCD as a renormalizable quantum field theory. In practice it is challenging to determine these in detail -- both theoretically and experimentally -- but the effort itself is triggering much progress on both sides. For example, it has been realized in recent years that many features of the QGP in the soft sector can be well described by relativistic fluid dynamics with relatively small shear and bulk viscosities \cite{Teaney:2009qa, Heinz:2013th, Busza:2018rrf}.

Here we investigate electrical conductivity, which has a similar status as the viscosities and thermal conductivity. We show that fluid dynamic considerations lead to a concrete expression for the electric current spectral density, which in turn determines the production rates of thermal photons and dileptons in the regime of small frequencies and momenta. Moreover, by using the space-time evolution of a fireball determined through relativistic fluid dynamics, we can show how thermal contributions to observable photon and dilepton spectra encode directly the information about electrical conductivity and an associated relaxation time in the soft regime. 

The connection between the spectral density, electrical conductivity and the photon and dilepton production rate is well known \cite{kadanoff_hydrodynamic_1963}. But so far there are hardly any calculations that connect electrical conductivity and measurements of dilepton or photon spectra directly.

Theoretical investigations of the electric current spectral function in the framework of thermal quantum field theory have a long history. Systematic analysis is possible in particular in the regime of perturbative QCD calculations at weak coupling and large temperatures (see refs.\ \cite{laine_thermal_2013, ghiglieri_next--leading_2013, Laine:2013vma} for an overview), as well as in the non-perturbative regime using lattice QCD simulations. The latter method addresses Euclidean times, however, and a numerically ill-defined analytic continuation is needed to go from there to real-time quantities of experimental relevance (see refs.\ \cite{ghiglieri_lattice_2016, Ce:2020tmx, Jackson:2019yao} for an overview). Unfortunately, neither perturbative nor lattice QCD calculations can be directly applied in the fluid dynamic regime of small frequencies and momenta \cite{ghiglieri_lattice_2016}.  

In the high temperature limit of QCD, the electrical conductivity can be calculated to leading log approximation within an effective kinetic theory. This was done by Arnold, Moore and Yaffe \cite{arnold_transport_2000,arnold_transport_2003} including quarks and leptons. Excluding the leptons leads to an electrical conductivity in the range of $0.19 <\sigma/T<2$ \cite{moritz_greif_electrical_2014}. 

Lattice QCD calculations using a fit to a specific ansatz for the spectral function done in the Euclidean domain led to values of the electrical conductivity in the range $0.003 <\sigma/T< 0.018$ \cite{aarts_electrical_2020}, see also ref.\ \cite{brandt_estimate_2018} for related work.

On the perturbative side, a complete leading-order result of the photon production rate was given in \cite{Arnold:2001ms} and extended to dileptons in \cite{Aurenche:2002wq}. Extensions to next-to-leading order were developed by Ghiglieri \textit{et al}.\ \cite{ghiglieri_next--leading_2013}, and by Laine \cite{Laine:2013vma}.
In the limit of high frequencies $\omega\gg T\gg M$, the spectral function was determined by Carrington \textit{et al}.\ in almost completely analytical form \cite{carrington_energetic_2008}.  including Compton scattering, annihilation processes and the Drell Yan process. 

Greif and Greiner discussed how to calculate the electrical conductivity based on the relaxation time approximation \cite{Greif:2016skc}. They found values of electrical conductivity in the range $0.001<\sigma/T<0.3$.

Yin derived constraints on the electrical conductivity from direct-photon spectra measured by the PHENIX collaboration in Au-Au collisions at $\sqrt{s_\text{NN}} = 200\,\mathrm{GeV}$ \cite{yin_electrical_2014,PHENIX:2014nkk}. For the effective electrical conductivity averaged over the space-time history of the fireball he obtained $0.04 < \sigma/T < 0.1$. 

In the following we use natural units $\hbar = c = k_B = \epsilon_0 = 1$ and the metric signature $(-,+,+,+)$, leading to a fluid velocity normalized by $u_\mu u^\mu=-1$.

\section{Electric current spectral function in the fluid dynamic regime}
\label{conductivity_hydro}
As a result of the electromagnetic U$(1)$ gauge symmetry, the electric current $J^\mu=(n, \mathbf{j})$ is conserved, 
\begin{equation}
\nabla_\mu J^\mu=0.
\label{eq:CurrentConservation}
\end{equation}
In the fluid dynamic regime of small frequencies and wave numbers, the space-time evolution of the electric current is determined by this conservation law, together with a constitutive relation that can be written in terms of a derivative expansion. For the latter we take
\begin{equation}
\label{eq:ConstitutiveCurrent}
    J^\alpha + \tau \Delta^\alpha_{\;\;\beta} u^\mu \nabla_\mu J^\beta =n u^\alpha+ \sigma \Delta^{\alpha\nu}E_\nu - D \Delta^{\alpha\nu} \partial_\nu n.
\end{equation}
The first terms on the left as well as on the right correspond to the leading order and contain no derivatives. The second and third term on the right-hand side are of first order in derivatives and involve the projector orthogonal to the fluid velocity $\Delta^{\alpha\beta}=g^{\alpha\beta}+u^\alpha u^\beta$.  We also use the electric field $E^\nu = \Delta^{\nu\mu}u^\rho F_{\mu\rho}$, where $F_{\mu\nu}=\partial_\mu A_\nu - \partial_\nu A_\mu$ is the electric field strength tensor, and $\sigma$ is the electrical conductivity. The last term in Eq.~\eqref{eq:ConstitutiveCurrent} realizes Fick's law with diffusion coefficient $D$.
The second term on the left-hand side in Eq.~\eqref{eq:ConstitutiveCurrent}, which is formally of second order in the derivative expansion, is introduced to ensure relativistic causality. The relaxation time $\tau$ corresponds to the time scale over which the current relaxes towards the local form of Ohm's and Fick's law described by the rest of the equation. Note that with this term, Eq.\ \eqref{eq:ConstitutiveCurrent} becomes an evolution law for the components of the current that are orthogonal to the fluid velocity.

One may now use linear response theory to determine different two-point correlation functions of electromagnetic current $J^\mu(x)$, see supplemental material. We are particularly interested in the spectral density $\rho(\omega, \mathbf{p}) = \rho^\mu_{\;\;\mu}(\omega, \mathbf{p})$ where $\rho^{\mu \nu}(\omega, \mathbf{p})=\text{Im}\, G_R^{\mu\nu}(\omega, \mathbf{p})$ is given by the imaginary part of the retarded correlator. In the fluid rest frame we find
\begin{equation}
\label{eq:rhoFluidApproximation}
    \rho(\omega, \mathbf{p}) = \frac{\sigma\omega(\omega^2-\mathbf{p}^2)}{(\tau \omega^2 - D \mathbf{p}^2)^2+\omega^2} + 2 \frac{\sigma\omega}{\tau^2 \omega^2+1}.
\end{equation}
The diffusion coefficient $D$ and the electrical conductivity $\sigma$ are related through the static charge susceptibility $\chi = (\partial n/ \partial \mu)|_T$ , as $D\chi=\sigma$.

From the concrete expression \eqref{eq:rhoFluidApproximation} one can obtain the Kubo relation for electrical conductivity in the low frequency limit of the spectral density on the light cone, or at vanishing spatial momentum,
\begin{equation}
\label{Kubo}
    \sigma=\frac{1}{2}\lim\limits_{\omega\rightarrow 0}\frac{1}{\omega} \rho(\omega,\mathbf{p}){\big |}_{\mathbf{p}^2=\omega^2} = \frac{1}{3}\lim\limits_{\omega\rightarrow 0}\frac{1}{\omega} \rho(\omega,\mathbf{p}){\big |}_{\mathbf{p}=0}.
\end{equation}
The combination $\rho(\omega, \mathbf{p})/\omega$ has a transport peak at small frequencies, the magnitude of which is determined by electrical conductivity. The width of this transport peak is determined by the inverse relaxation time $\tau^{-1}$. Quite generally, we expect Eq.\ \eqref{eq:rhoFluidApproximation} to capture the behavior of the spectral density in the hydrodynamic regime of small frequencies and momenta compared to the temperature scale, $\omega,|\vec{p}|\ll T$.

Let us also note here that the dispersion relation derived in the supplemental material has two transverse modes with $\omega=-i/\tau$ and accordingly vanishing asymptotic group velocity, $\lim_{|\mathbf{p}|\to\infty}(d\omega/dp)=0$. In contrast, the longitudinal mode has $\lim_{|\mathbf{p}|\to\infty}(d\omega/dp)=\sqrt{D/\tau}$ and relativistic causality requires therefore 
\begin{equation}
\label{eq:causalityBound}
    \tau>D = \frac{\sigma}{\chi}.
\end{equation}

From dimensional analysis one infers that $\sigma = \hat \sigma T$ with dimensionless $\hat\sigma = \sigma/T$. In a similar way one can write $D=\hat \sigma / (\hat \chi T)$ and $\tau=\hat \tau/T$. For our numerical calculations we will assume that $\hat \sigma$, $\hat \chi$ and $\hat \tau$ are independent of temperature. This should be a realistic approximation for QCD at large enough temperatures and is supported by previous work, see for example refs.\ \cite{Greif:2016skc, banerjee_heavy_2012}. The charge susceptibility $\hat \chi$ has been determined through lattice QCD simulations \cite{Borsanyi:2011sw, bazavov_fluctuations_2012}. For large temperatures one finds $\hat{\chi}\approx 0.6$ and we will use this value throughout. 

\begin{figure}[t]
	\centering 
	\includegraphics[width=1\linewidth]{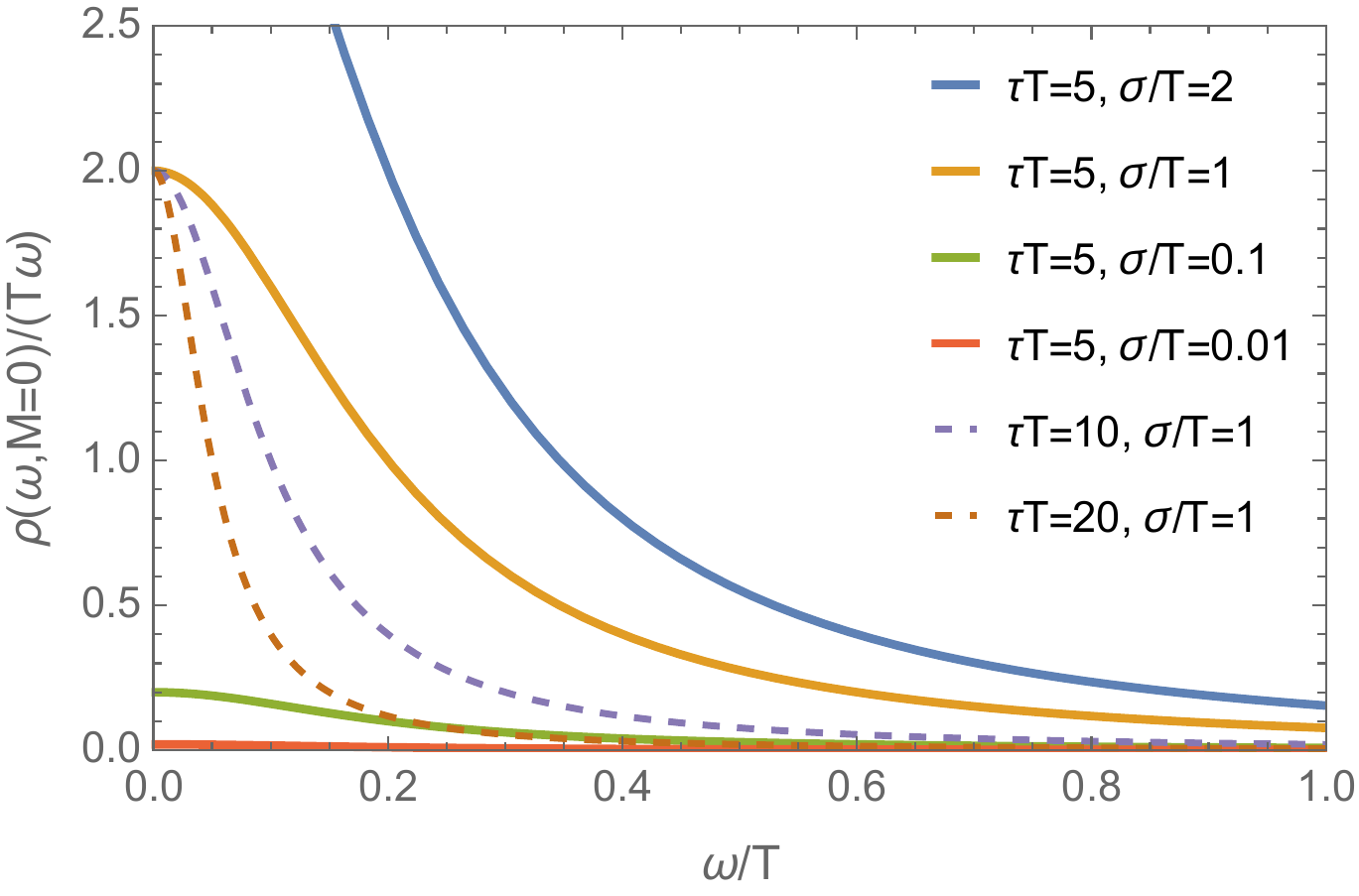}
	\caption{Spectral density for the electric current derived from fluid dynamic considerations as given by Eq.~\eqref{eq:rhoFluidApproximation} at vanishing invariant mass, $M=0$. Dashed lines illustrate the dependence on the relaxation time, solid lines show the dependence on conductivity. Electric conductivity is proportional to the height of the spectral peak, whereas an increasing relaxation time makes it more narrow.}
	\label{fig:dependence_spectral_density}
\end{figure}

In Fig.\ \ref{fig:dependence_spectral_density} we show the spectral density in Eq.\ \eqref{eq:rhoFluidApproximation} as a function of frequency $\omega$ for different values of conductivity $\sigma/T$ and relaxation time $\tau T$ for vanishing invariant mass (real photons).

Let us remark here that we are neglecting any possible modification of the spectral density or conductivity due to magnetic fields, see refs.\ \cite{nam_electrical_2012,hattori_electrical_2016, feng_electric_2017} for related discussions.

\section{Photon and dilepton production rates}
The photon production rate per unit volume and time in (local) thermal equilibrium is given by
\begin{align}
\label{photonproductionrate}
    p^0\frac{d R}{d^3p}&=\frac{1}{(2\pi)^3}n_\text{B}(\omega)\rho(\omega), 
\end{align} 
with the Bose-Einstein distribution factor $n_\text{B}(\omega)=(e^{\omega/T}-1)^{-1}$ and the frequency in the fluid rest frame $\omega = -u_\mu p^\mu$. The spectral density $\rho(\omega)$ is here evaluated on the mass shell for photons, i.e., at vanishing invariant mass, $M^2=-p^\mu p_\mu=0$, and it is also a function of temperature $T$.

In a similar way the thermal dilepton production rate per unit volume and time can be calculated to be
\begin{align}
\label{dileptonnproductionrate}
    \frac{dR}{d^4p}   =&\frac{\alpha}{12\pi^4}\frac{1}{M^2} n_\text{B}(\omega) \, \rho(\omega,M)\\& \times\left(1+\frac{2m^2}{M^2}\right)\sqrt{1-\frac{4m^2}{M^2}}\Theta(M^2-4m^2),\nonumber
\end{align}
with the momentum of the dilepton pair being $p^\mu=p^\mu_1+p^\mu_2$, $m$ the lepton mass, and $\alpha=e^2/(4\pi)$ the electromagnetic fine structure constant.

Combining Eqs.~\eqref{Kubo}, \eqref{photonproductionrate} and \eqref{dileptonnproductionrate}, it is expected that production rates reach a finite value proportional to electrical conductivity in the limit of small energies. Using Bjorken coordinates in momentum space, this schematically reads
\begin{equation}
\label{limit}
    \lim\limits_{p_T\rightarrow0}\frac{dN_\gamma}{p_T dp_T d\eta}\propto\hat{\sigma}.
\end{equation}
The proportionality factor contains an integral over the space-time volume occupied by the quark-gluon plasma and depends on temperature and fluid velocity of the latter. 
Eq.~\eqref{limit} provides the basis of our further analysis which aims to make this relation more precise. We want to investigate how it is possible to get information about electrical conductivity from direct measurements of the soft photon and dilepton spectra at particle colliders.

\section{Photon and dilepton spectra}
\label{results}
In the following we will determine photon and dilepton spectra by integrating Eqs.\ \eqref{photonproductionrate} and \eqref{dileptonnproductionrate} over the space-time volume of a heavy-ion collision. In order to obtain the local temperature $T(x)$ and fluid velocity $u^\mu(x)$ we use the code \textsc{FluiduM} \cite{floerchinger_fluid_2019}, with parameters similar to those described in refs.\ \cite{Devetak:2019lsk, Bernhard:2016tnd}. Our calculations address central (within the $0 - 5\%$ class) Pb-Pb collisions with $\sqrt{s_\text{NN}}= 5.02$ TeV. Integration is performed up to a freeze-out hypersurface corresponding to $T=140$ MeV. This implies that we consider only particle production from the QGP itself, and not from a subsequent hadron resonance gas (HRG) phase. According to ref.\ \cite{Turbide:2003si}, both contributions account for about equal proportion of the full photon and dilepton yields. We leave the contribution from the HRG phase for future work, but note that the methods we develop could also be applied there, to the extent that fluid dynamics is a good approximation.

For the spectral density we use the result obtained for the fluid limit in Eq.~\eqref{eq:rhoFluidApproximation}, as applicable for small frequencies and momenta. 

For definiteness, we concentrate on four values of the electrical conductivity to temperature ratio, $\sigma/T={0.01,\,0.1,\,1,\,2}$. We fix the relaxation time to be $\tau=5/T$ such that the causality bound in Eq.~\eqref{eq:causalityBound} is satisfied for all choices of $\sigma/T$. Other possible values should be investigated in the future. 

\begin{figure}[t]
	\centering
	\includegraphics[width=1\linewidth]{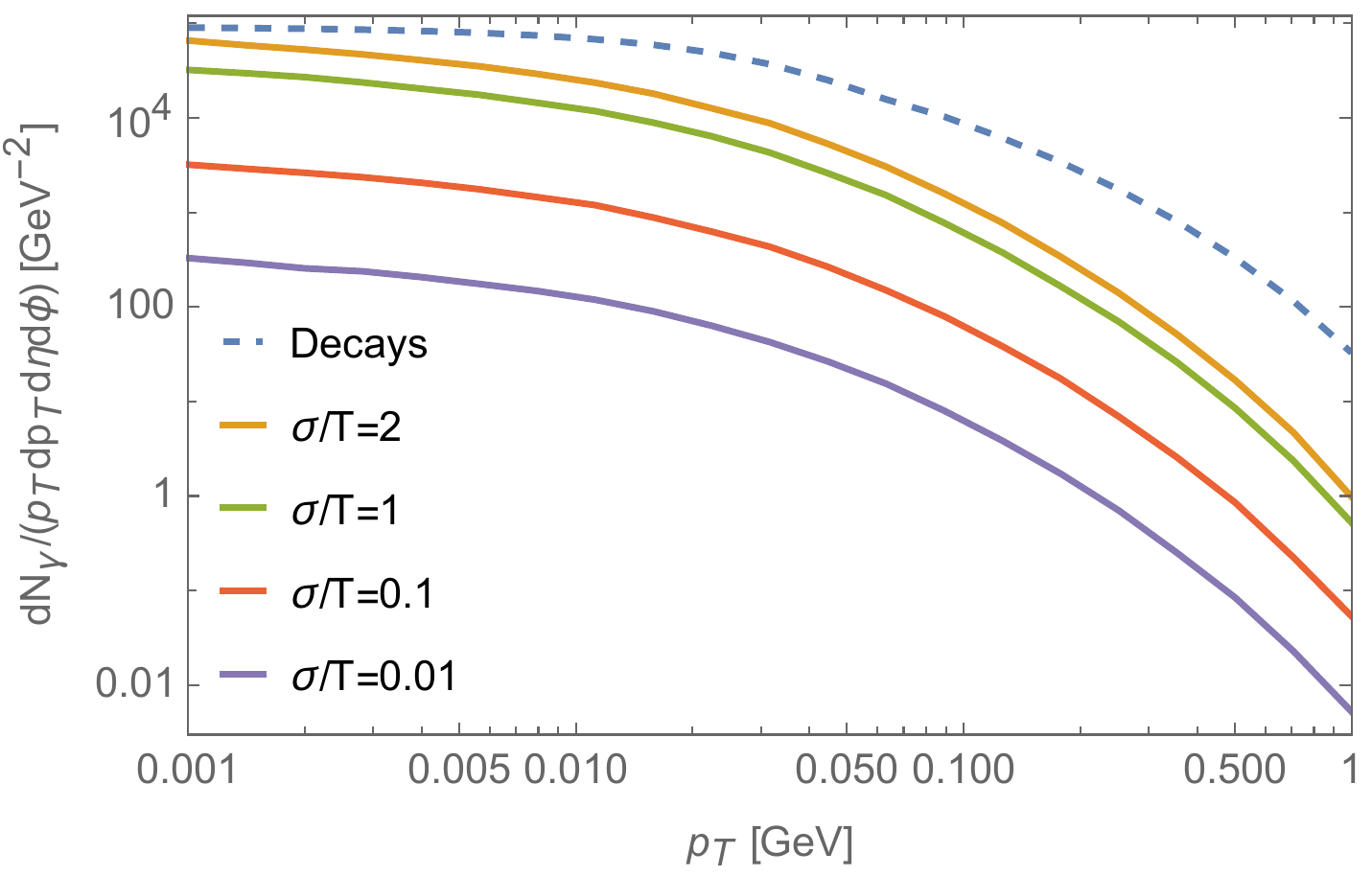}
	\caption{Thermal photon transverse momentum spectrum computed by integrating the fluid dynamic spectral density \eqref{eq:rhoFluidApproximation} over the space-time volume of a fireball using the fluid dynamics code \textsc{FluiduM}, for different values of electrical conductivity (colored). We choose here the relaxation time $\tau=5/T$. The black curves shows the spectrum of decay photons from the decay $\pi^0\to\gamma\gamma$ computed calculated with the methods of ref.\ \cite{mazeliauskas_fast_2019}.}
	\label{fig:photon}
\end{figure}

In Fig. \ref{fig:photon} we show the resulting transverse momentum spectrum for photons (which is independent of azimuthal angle $\phi$) for different values of the electrical conductivity (colored lines). It should be emphasized again that the assumptions underlying the fluid dynamic description become questionable at large momenta, albeit the blue shift caused by the radial expansion can boost thermal photons that are soft in the fluid rest frame to somewhat larger momenta in the laboratory frame.

\begin{figure}[t]
	\centering
	\includegraphics[width=1\linewidth]{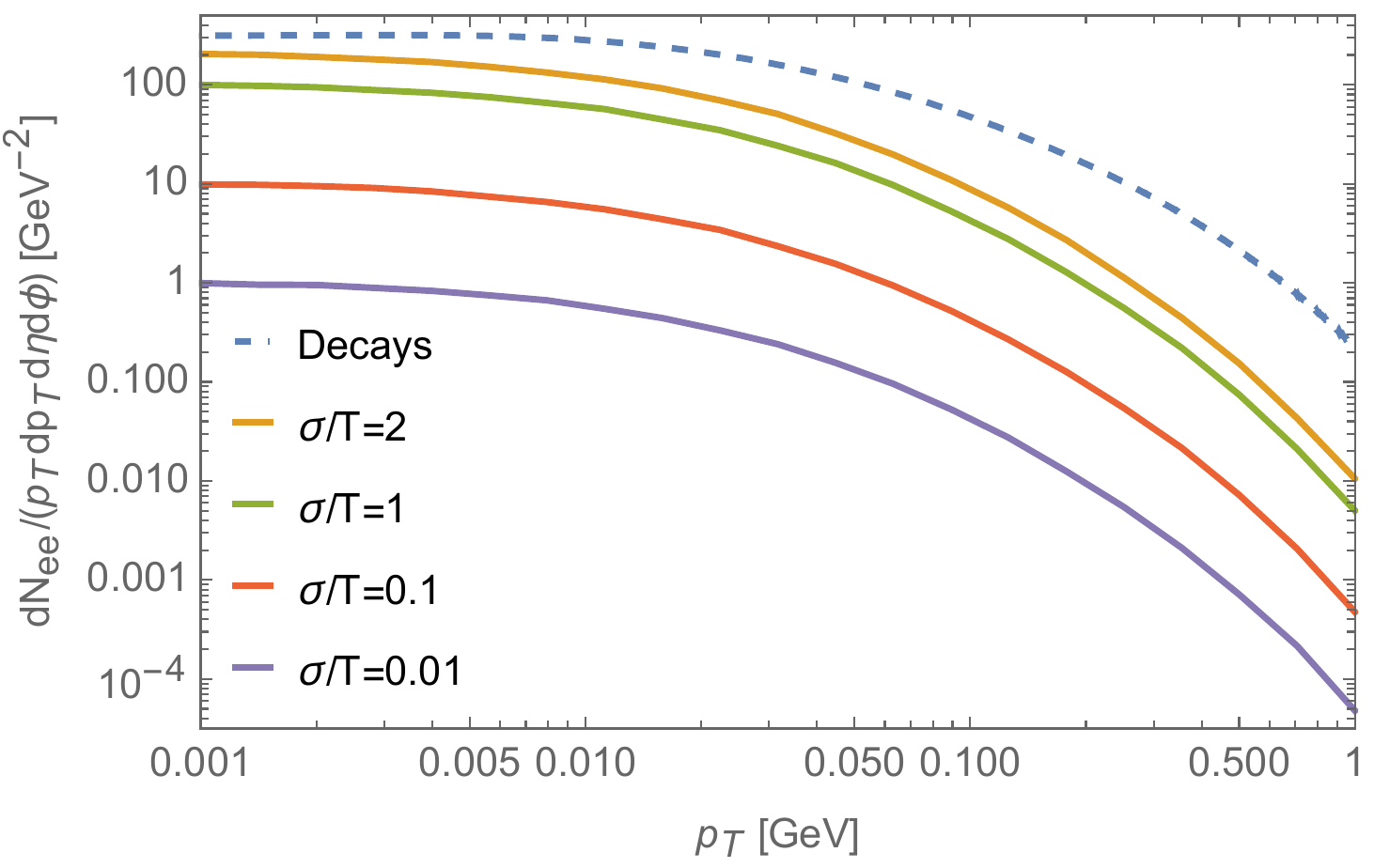}
	\caption{Thermal dielectron spectrum as a function of the pair transverse momentum for different values of the electrical conductivity and $\tau=5/T$. The invariant mass has here been integrated in the kinematically allowed region. 
	We compare the results with computations of dielectron pairs from decays of short-lived particles computed with a Monte Carlo simulation (black line).}
	\label{fig:dielectron_pt}
\end{figure}

\begin{figure}[t]
	\centering
	\includegraphics[width=1\linewidth]{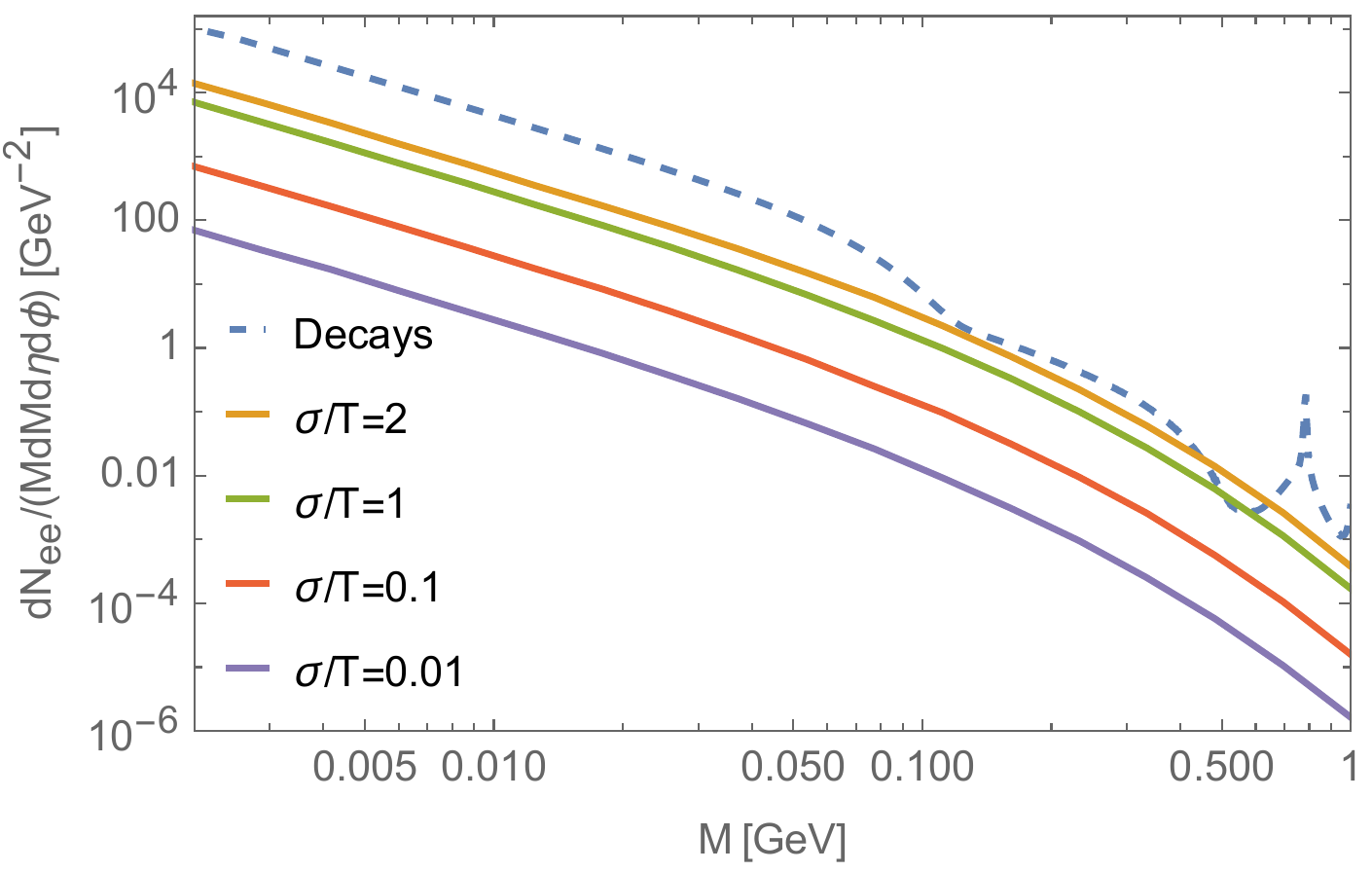}
	\caption{Thermal dielectron spectrum similar to fig.\ \ref{fig:dielectron_pt}, but now as a function of invariant mass $M$ and integrated with respect to the pair transverse momentum. The curves are only shown in the kinematically allowed region $M>2 m_e$. We also show the spectrum of dielectrons produced by decays of short-lived hadrons for comparison.}
	\label{fig:dielectron_m}
\end{figure}

In Fig.\ \ref{fig:dielectron_pt} we show spectra of dielectrons as a function of transverse momentum based on the integration of Eq.~\eqref{dileptonnproductionrate} with the spectral density in Eq.~\eqref{eq:rhoFluidApproximation} for different values of $\sigma/T$. We have integrated over the invariant mass in the kinematically allowed range $M>2 m_e$. Similarly, in Fig.\ \ref{fig:dielectron_m} we show the dielectron spectrum as a function of the invariant mass $M$, now integrated with respect to the pair transverse momentum. One observes that the electrical conductivity has a substantial influence in all cases. 

\section{Experimental access to the electrical conductivity}

The experimental access to the electrical conductivity is challenging as photons and dielectrons produced in the decays of short-lived hadrons dominate the yields at low momenta and low masses. This is illustrated in Figs.~\ref{fig:photon}, \ref{fig:dielectron_pt} and \ref{fig:dielectron_m} which show the background from hadron decays in addition to the thermal yields. The largest contribution to the background comes from the decays of neutral pions and $\eta$ mesons. The decay photon background was calculated with the FastReso \cite{mazeliauskas_fast_2019} code based on the hadron spectra from FluiduM. The dielectron background from hadron decays was calculated in a Monte Carlo approach using PYTHIA~8 particle decay routines \cite{Sjostrand:2007gs}. Only contributions from light-flavor hadrons including $\pi^0$, $\eta$, $\omega$, $\rho^0$, $\eta'$, and $\phi$ were considered. The neutral pion spectrum from FluiduM/FastReso was taken as input and $m_T$ scaling \cite{WA80:1995whm} was assumed for the spectrum of the other  mesons. For photons, the signal-to-background ratio at $p_T=1$ MeV varies from about 70\% to 0.4\% for conductivities in the range $\sigma/T = 2 \text{--}0.01$. For the same range of conductivities, the signal-to-background ratios of the dielectrons at low $p_T$ varies from about 67\% to about 0.3\%. Assuming experimental uncertainties of neutral pion and $\eta$ meson yields at low $p_T$ of a few percent or more, the challenge of measuring small values of the electrical conductivity becomes apparent.

In the case of dielectrons, a promising approach is to consider only dielectrons at low $p_T$ above a minimum mass. The mass distribution of electron pairs from $\pi^0$ and $\eta$ Dalitz decays is given by the Kroll-Wada formula \cite{Kroll:1955zu,Landsberg:1985gaz} and follows the same $1/M^2$ form as the thermal yields over a wide mass range. Accepting only dielectrons with $M \gtrsim 100~\mathrm{MeV}$ greatly reduces the background as dielectrons from $\pi^0$ Dalitz decays quickly fall off close to the neutral pion mass.
For instance, the signal-to-background ratio at low $p_T$ for $M > 100~\mathrm{MeV}$ for the case of $\sigma/T = 0.1$ improves by factor of about 4 to about 6.7\%.

In the case of photons, Hanbury Brown-Twiss (HBT) correlations provide an attractive way to determine photon yields a low $p_T$ \cite{WA98:2003ukc,Peressounko:2003cf}. Assuming a fully chaotic source, Bose-Einstein correlations of photons produced in the QGP give rise to an enhanced production of photon pairs with small invariant relative momenta $Q_\mathrm{inv}$. Experimentally, all photons need to be considered, including those from decays of hadrons. The latter essentially dilute the observed correlation strength of the QGP photons, making the measured correlation strength a measure of the ratio $N_\mathrm{\gamma,dir}/N_\mathrm{\gamma,total}$ of direct photons to all photons. Photon pairs from neutral pion or $\eta$ meson decays produce a peak in the correlation function at higher $Q_\mathrm{inv}$ values. The photon HBT method thus allows one to measure direct photons yields at low $p_T$ without a statistical subtraction of decay photons from neutral pions and $\eta$ mesons. At CERN SPS energies, photon yields obtained through photon HBT were found to be in qualitative agreement with theoretical calculations for the electrical conductivity of a meson gas \cite{Fernandez-Fraile:2009eug}.   

\section{Conclusions}

In conclusion, the electric current spectral function can be determined in the regime of small frequencies and momenta from fluid dynamic considerations. It is parameterized by the electrical conductivity, the static electric susceptibility, and the relaxation time for the electric current. An integral of this quantity over the space-time history of a heavy-ion collision event, which we have calculated, is accessible experimentally, through thermal photon and dilepton spectra. Advances in detector technology may soon allow one to resolve the latter in the relevant regime of small transverse momenta $p_T \ll T$ and invariant mass $M \ll T$ \cite{Adamova:2019vkf}. A confounding factor are background photons and dielectrons from decays of final-state hadrons. For dielectrons, the signal-to-background ratio can be significantly improved if only pairs above a minimum mass are considered. For the measurement of direct photons at small momenta, the HBT method is suitable, since this method is insensitive to photons from the decay of hadrons.

\begin{acknowledgments}
This work is supported by the DFG (German Research Foundation) -- Project-ID 273811115 -- SFB 1225 ISOQUANT and under Germany's Excellence Strategy EXC 2181/1 - 390900948 (the Heidelberg STRUCTURES Excellence Cluster) as well as FL 736/3-1.

\end{acknowledgments}

\section*{Supplemental Material: Electrical conductivity and spectral density}
\label{appendix1}
In this appendix we discuss linear response theory for the electromagnetic current in the low frequency and momentum, or hydrodynamic limit.

We start from an interaction Hamiltonian with an external perturbation in the electromagnetic gauge field 
$\Delta H(t) = - \int_{\mathbf{x}} A_\mu(t, \mathbf{x}) J^\mu(t, \mathbf{x})$
Here $j^0 = n$ is the charge density of the quark-gluon plasma and $\mathbf{J}$ is the electric current density. The electric field is given as usual by $\mathbf{E}=-\partial_t\mathbf{A}+\mathbf{\nabla} A_0$.

We define the retarded response function (following the conventions of ref.\ \cite{floerchinger_variational_2016}) as
\begin{equation}
    G_R^{\mu\nu}(x-y)=i\theta(x^0-y^0)\langle [J^\mu(x),J^\nu(y)]\rangle.
\end{equation}
Consequently, the response of the current to a perturbation in the gauge field is
\begin{equation}
\label{Gdefin}
    \delta\langle J^\mu(x)\rangle=\int_y G_R^{\mu\nu}(x-y) \,\delta A_\nu(y).
\end{equation}

In the following we want to obtain the retarded correlation function defined through Eq.~\eqref{Gdefin} in the fluid dynamic limit corresponding to large times and distances, or small frequencies and momenta. We are using to that end the constitutive relation \eqref{eq:ConstitutiveCurrent}.

In a static situation, and in the fluid rest frame, Eq.~\eqref{eq:ConstitutiveCurrent} gives for constant temperature
\begin{equation}
\label{eq:currentOhmLaw}
\mathbf{J} =\sigma\mathbf{E}- D\mathbf{\nabla} n = \sigma \mathbf{\nabla} A_0 - D \chi \mathbf{\nabla}\mu. 
\end{equation}
We use for the second equality the relation $dn = \chi d\mu$ with static charge susceptibility $\chi = (\partial n / \partial \mu)|_T$. Because $A^0=-A_0$ and $\mu$ appear in the microscopic action in the Euclidean Matsubara formulation only in the combination $A^0+\mu$, one infers Einsteins relation $D\chi=\sigma$.

The charge conservation law \eqref{eq:CurrentConservation} together with Eq.~\eqref{eq:ConstitutiveCurrent} give equations of motions for all components of the current $J^\mu$. In the following we use these to infer the retarded correlation function $G_R^{\mu\nu}$. 

Considering first a vanishing gauge field, $A_\nu=0$, the equations of motion can be written in the following matrix form, where we allowed for two (possibly frequency and momentum dependent) normalization factors $N_1$ and $N_2$,
\begin{equation}
    \underbrace{\begin{pmatrix}
    -iN_1\omega & iN_1\mathbf{p} \\
    iN_2 D\mathbf{p} & N_2(1-i\omega\tau)\mathbbm{1}
    \end{pmatrix}}_{P(\omega,\mathbf{p})} J(\omega,\mathbf{p})=0.
\end{equation}
The corresponding retarded Greens function is given by $G_R(\omega,\mathbf{p})=P^{-1}(\omega,\mathbf{p})$. 
The factors $N_1$ and $N_2$ can be determined in the limit of small frequencies and momenta. Specifically, for $\mathbf{p}=0$ and $\omega\to 0$ one should have $N_1 = 1/(i\chi \omega)$ to fulfill $dn=\chi dA^0$ and $N_2=1/(i\sigma \omega)$ to fulfill Eq.~\eqref{eq:currentOhmLaw}. With this we find for the inverse retarded propagator
\begin{equation}
\label{correlator}
    (G_R^{-1})_{\mu\nu}=P_{\mu\nu} = \begin{pmatrix}
    \frac{-1}{\chi} & \frac{\mathbf{p}}{\chi \omega}\\
    \frac{\mathbf{p}}{\chi \omega} & \left[\frac{-i}{\sigma\omega}-\frac{\tau}{\sigma}\right]\mathbbm{1}
    \end{pmatrix}.
\end{equation}
Matrix inversion leads to
\begin{equation}
\label{eq:RetardedPropagatorFull}
    G_R^{\mu\nu} = \begin{pmatrix} \mathsf{A} && \mathsf{B} \frac{\mathbf{p}}{\omega} \\ \mathsf{B} \frac{\mathbf{p}}{\omega} && \mathsf{C} \delta_{mn} + \mathsf{D} \frac{p_m p_n}{\mathbf{p}^2} \end{pmatrix},
\end{equation}
with the coefficients
\begin{equation}
\begin{split}
    & \mathsf{A} = \tfrac{\mathbf{p}^2}{\omega^2} \mathsf{B}-\chi, \quad\quad\quad\quad\; \mathsf{B} = \left[ \tfrac{\mathbf{p}^2}{\chi \omega^2}-\tfrac{i}{\sigma \omega} - \tfrac{\tau}{\sigma} \right]^{-1}, \\
    & \mathsf{C} = \left[ -\tfrac{i}{\sigma\omega} - \tfrac{\tau}{\sigma} \right]^{-1}, \quad\quad \mathsf{D} = \mathsf{B}- \mathsf{C}.
\end{split}
\end{equation}

The spectral function is given by the imaginary part of the retarded propagator $\rho^{\mu \nu}=\text{Im}\, G_R^{\mu\nu}$. From \eqref{eq:RetardedPropagatorFull} one finds 
\begin{equation}
       \rho^{\mu\nu} = \begin{pmatrix} \hat{\mathsf{A}} && \hat{\mathsf{B}} \frac{\mathbf{p}}{\omega} \\ \hat{\mathsf{B}} \frac{\mathbf{p}}{\omega} && \hat{\mathsf{C}} \delta_{mn} + \hat{\mathsf{D}} \frac{p_m p_n}{\mathbf{p}^2} \end{pmatrix},
\end{equation}
with coefficients
\begin{equation}
\begin{split}
    & \hat{\mathsf{A}} = \tfrac{\mathbf{p}^2}{\omega^2} \hat{\mathsf{B}}, \quad\quad\quad\quad\quad \hat{\mathsf{B}} = \frac{\sigma\omega^3}{(\tau \omega^2 - D \mathbf{p}^2)^2+\omega^2}, \\
    & \hat{\mathsf{C}} = \frac{\sigma\omega}{\tau^2 \omega^2+1},\quad\quad\;\;\; \hat{\mathsf{D}} = \hat{\mathsf{B}}- \hat{\mathsf{C}}.
\end{split}
\end{equation}
The trace $\rho=\rho_{\;\mu}^\mu$ evaluates to the expression given in Eq.\ \eqref{eq:rhoFluidApproximation}.

\bibliography{Bib}

\begin{thebibliography}{41}%
\makeatletter
\providecommand \@ifxundefined [1]{%
 \@ifx{#1\undefined}
}%
\providecommand \@ifnum [1]{%
 \ifnum #1\expandafter \@firstoftwo
 \else \expandafter \@secondoftwo
 \fi
}%
\providecommand \@ifx [1]{%
 \ifx #1\expandafter \@firstoftwo
 \else \expandafter \@secondoftwo
 \fi
}%
\providecommand \natexlab [1]{#1}%
\providecommand \enquote  [1]{``#1''}%
\providecommand \bibnamefont  [1]{#1}%
\providecommand \bibfnamefont [1]{#1}%
\providecommand \citenamefont [1]{#1}%
\providecommand \href@noop [0]{\@secondoftwo}%
\providecommand \href [0]{\begingroup \@sanitize@url \@href}%
\providecommand \@href[1]{\@@startlink{#1}\@@href}%
\providecommand \@@href[1]{\endgroup#1\@@endlink}%
\providecommand \@sanitize@url [0]{\catcode `\\12\catcode `\$12\catcode
  `\&12\catcode `\#12\catcode `\^12\catcode `\_12\catcode `\%12\relax}%
\providecommand \@@startlink[1]{}%
\providecommand \@@endlink[0]{}%
\providecommand \url  [0]{\begingroup\@sanitize@url \@url }%
\providecommand \@url [1]{\endgroup\@href {#1}{\urlprefix }}%
\providecommand \urlprefix  [0]{URL }%
\providecommand \Eprint [0]{\href }%
\providecommand \doibase [0]{https://doi.org/}%
\providecommand \selectlanguage [0]{\@gobble}%
\providecommand \bibinfo  [0]{\@secondoftwo}%
\providecommand \bibfield  [0]{\@secondoftwo}%
\providecommand \translation [1]{[#1]}%
\providecommand \BibitemOpen [0]{}%
\providecommand \bibitemStop [0]{}%
\providecommand \bibitemNoStop [0]{.\EOS\space}%
\providecommand \EOS [0]{\spacefactor3000\relax}%
\providecommand \BibitemShut  [1]{\csname bibitem#1\endcsname}%
\let\auto@bib@innerbib\@empty
\bibitem [{\citenamefont {Teaney}(2010)}]{Teaney:2009qa}%
  \BibitemOpen
  \bibfield  {author} {\bibinfo {author} {\bibfnamefont {D.~A.}\ \bibnamefont
  {Teaney}},\ }\bibinfo {title} {{Viscous Hydrodynamics and the Quark Gluon
  Plasma}},\ in\ \href {https://doi.org/10.1142/9789814293297_0004} {\emph
  {\bibinfo {booktitle} {{Quark-gluon plasma 4}}}},\ \bibinfo {editor} {edited
  by\ \bibinfo {editor} {\bibfnamefont {R.~C.}\ \bibnamefont {Hwa}}\ and\
  \bibinfo {editor} {\bibfnamefont {X.-N.}\ \bibnamefont {Wang}}}\ (\bibinfo
  {year} {2010})\ pp.\ \bibinfo {pages} {207--266},\ \Eprint
  {https://arxiv.org/abs/0905.2433} {arXiv:0905.2433 [nucl-th]} \BibitemShut
  {NoStop}%
\bibitem [{\citenamefont {Heinz}\ and\ \citenamefont
  {Snellings}(2013)}]{Heinz:2013th}%
  \BibitemOpen
  \bibfield  {author} {\bibinfo {author} {\bibfnamefont {U.}~\bibnamefont
  {Heinz}}\ and\ \bibinfo {author} {\bibfnamefont {R.}~\bibnamefont
  {Snellings}},\ }\bibfield  {title} {\bibinfo {title} {{Collective flow and
  viscosity in relativistic heavy-ion collisions}},\ }\href
  {https://doi.org/10.1146/annurev-nucl-102212-170540} {\bibfield  {journal}
  {\bibinfo  {journal} {Ann. Rev. Nucl. Part. Sci.}\ }\textbf {\bibinfo
  {volume} {63}},\ \bibinfo {pages} {123} (\bibinfo {year} {2013})},\ \Eprint
  {https://arxiv.org/abs/1301.2826} {arXiv:1301.2826 [nucl-th]} \BibitemShut
  {NoStop}%
\bibitem [{\citenamefont {Busza}\ \emph {et~al.}(2018)\citenamefont {Busza},
  \citenamefont {Rajagopal},\ and\ \citenamefont {van~der
  Schee}}]{Busza:2018rrf}%
  \BibitemOpen
  \bibfield  {author} {\bibinfo {author} {\bibfnamefont {W.}~\bibnamefont
  {Busza}}, \bibinfo {author} {\bibfnamefont {K.}~\bibnamefont {Rajagopal}},\
  and\ \bibinfo {author} {\bibfnamefont {W.}~\bibnamefont {van~der Schee}},\
  }\bibfield  {title} {\bibinfo {title} {{Heavy Ion Collisions: The Big
  Picture, and the Big Questions}},\ }\href
  {https://doi.org/10.1146/annurev-nucl-101917-020852} {\bibfield  {journal}
  {\bibinfo  {journal} {Ann. Rev. Nucl. Part. Sci.}\ }\textbf {\bibinfo
  {volume} {68}},\ \bibinfo {pages} {339} (\bibinfo {year} {2018})},\ \Eprint
  {https://arxiv.org/abs/1802.04801} {arXiv:1802.04801 [hep-ph]} \BibitemShut
  {NoStop}%
\bibitem [{\citenamefont {Kadanoff}\ and\ \citenamefont
  {Martin}(1963)}]{kadanoff_hydrodynamic_1963}%
  \BibitemOpen
  \bibfield  {author} {\bibinfo {author} {\bibfnamefont {L.~P.}\ \bibnamefont
  {Kadanoff}}\ and\ \bibinfo {author} {\bibfnamefont {P.~C.}\ \bibnamefont
  {Martin}},\ }\bibfield  {title} {\bibinfo {title} {Hydrodynamic equations and
  correlation functions},\ }\href@noop {} {\bibfield  {journal} {\bibinfo
  {journal} {Annals of Physics}\ }\textbf {\bibinfo {volume} {24}},\ \bibinfo
  {pages} {419} (\bibinfo {year} {1963})}\BibitemShut {NoStop}%
\bibitem [{\citenamefont {Laine}(2013{\natexlab{a}})}]{laine_thermal_2013}%
  \BibitemOpen
  \bibfield  {author} {\bibinfo {author} {\bibfnamefont {M.}~\bibnamefont
  {Laine}},\ }\bibfield  {title} {\bibinfo {title} {{Thermal 2-loop master
  spectral function at finite momentum}},\ }\href
  {https://doi.org/10.1007/JHEP05(2013)083} {\bibfield  {journal} {\bibinfo
  {journal} {JHEP}\ }\textbf {\bibinfo {volume} {05}},\ \bibinfo {pages}
  {083}},\ \Eprint {https://arxiv.org/abs/1304.0202} {arXiv:1304.0202 [hep-ph]}
  \BibitemShut {NoStop}%
\bibitem [{\citenamefont {Ghiglieri}\ \emph {et~al.}(2013)\citenamefont
  {Ghiglieri}, \citenamefont {Hong}, \citenamefont {Kurkela}, \citenamefont
  {Lu}, \citenamefont {Moore},\ and\ \citenamefont
  {Teaney}}]{ghiglieri_next--leading_2013}%
  \BibitemOpen
  \bibfield  {author} {\bibinfo {author} {\bibfnamefont {J.}~\bibnamefont
  {Ghiglieri}}, \bibinfo {author} {\bibfnamefont {J.}~\bibnamefont {Hong}},
  \bibinfo {author} {\bibfnamefont {A.}~\bibnamefont {Kurkela}}, \bibinfo
  {author} {\bibfnamefont {E.}~\bibnamefont {Lu}}, \bibinfo {author}
  {\bibfnamefont {G.~D.}\ \bibnamefont {Moore}},\ and\ \bibinfo {author}
  {\bibfnamefont {D.}~\bibnamefont {Teaney}},\ }\bibfield  {title} {\bibinfo
  {title} {{Next-to-leading order thermal photon production in a weakly coupled
  quark-gluon plasma}},\ }\href {https://doi.org/10.1007/JHEP05(2013)010}
  {\bibfield  {journal} {\bibinfo  {journal} {JHEP}\ }\textbf {\bibinfo
  {volume} {05}},\ \bibinfo {pages} {010}},\ \Eprint
  {https://arxiv.org/abs/1302.5970} {arXiv:1302.5970 [hep-ph]} \BibitemShut
  {NoStop}%
\bibitem [{\citenamefont {Laine}(2013{\natexlab{b}})}]{Laine:2013vma}%
  \BibitemOpen
  \bibfield  {author} {\bibinfo {author} {\bibfnamefont {M.}~\bibnamefont
  {Laine}},\ }\bibfield  {title} {\bibinfo {title} {{NLO thermal dilepton rate
  at non-zero momentum}},\ }\href {https://doi.org/10.1007/JHEP11(2013)120}
  {\bibfield  {journal} {\bibinfo  {journal} {JHEP}\ }\textbf {\bibinfo
  {volume} {11}},\ \bibinfo {pages} {120}},\ \Eprint
  {https://arxiv.org/abs/1310.0164} {arXiv:1310.0164 [hep-ph]} \BibitemShut
  {NoStop}%
\bibitem [{\citenamefont {Ghiglieri}\ \emph {et~al.}(2016)\citenamefont
  {Ghiglieri}, \citenamefont {Kaczmarek}, \citenamefont {Laine},\ and\
  \citenamefont {Meyer}}]{ghiglieri_lattice_2016}%
  \BibitemOpen
  \bibfield  {author} {\bibinfo {author} {\bibfnamefont {J.}~\bibnamefont
  {Ghiglieri}}, \bibinfo {author} {\bibfnamefont {O.}~\bibnamefont
  {Kaczmarek}}, \bibinfo {author} {\bibfnamefont {M.}~\bibnamefont {Laine}},\
  and\ \bibinfo {author} {\bibfnamefont {F.}~\bibnamefont {Meyer}},\ }\bibfield
   {title} {\bibinfo {title} {{Lattice constraints on the thermal photon
  rate}},\ }\href {https://doi.org/10.1103/PhysRevD.94.016005} {\bibfield
  {journal} {\bibinfo  {journal} {Phys. Rev. D}\ }\textbf {\bibinfo {volume}
  {94}},\ \bibinfo {pages} {016005} (\bibinfo {year} {2016})},\ \Eprint
  {https://arxiv.org/abs/1604.07544} {arXiv:1604.07544 [hep-lat]} \BibitemShut
  {NoStop}%
\bibitem [{\citenamefont {C\`e}\ \emph {et~al.}(2020)\citenamefont {C\`e},
  \citenamefont {Harris}, \citenamefont {Meyer}, \citenamefont {Steinberg},\
  and\ \citenamefont {Toniato}}]{Ce:2020tmx}%
  \BibitemOpen
  \bibfield  {author} {\bibinfo {author} {\bibfnamefont {M.}~\bibnamefont
  {C\`e}}, \bibinfo {author} {\bibfnamefont {T.}~\bibnamefont {Harris}},
  \bibinfo {author} {\bibfnamefont {H.~B.}\ \bibnamefont {Meyer}}, \bibinfo
  {author} {\bibfnamefont {A.}~\bibnamefont {Steinberg}},\ and\ \bibinfo
  {author} {\bibfnamefont {A.}~\bibnamefont {Toniato}},\ }\bibfield  {title}
  {\bibinfo {title} {{Rate of photon production in the quark-gluon plasma from
  lattice QCD}},\ }\href {https://doi.org/10.1103/PhysRevD.102.091501}
  {\bibfield  {journal} {\bibinfo  {journal} {Phys. Rev. D}\ }\textbf {\bibinfo
  {volume} {102}},\ \bibinfo {pages} {091501} (\bibinfo {year} {2020})},\
  \Eprint {https://arxiv.org/abs/2001.03368} {arXiv:2001.03368 [hep-lat]}
  \BibitemShut {NoStop}%
\bibitem [{\citenamefont {Jackson}\ and\ \citenamefont
  {Laine}(2019)}]{Jackson:2019yao}%
  \BibitemOpen
  \bibfield  {author} {\bibinfo {author} {\bibfnamefont {G.}~\bibnamefont
  {Jackson}}\ and\ \bibinfo {author} {\bibfnamefont {M.}~\bibnamefont
  {Laine}},\ }\bibfield  {title} {\bibinfo {title} {{Testing thermal photon and
  dilepton rates}},\ }\href {https://doi.org/10.1007/JHEP11(2019)144}
  {\bibfield  {journal} {\bibinfo  {journal} {JHEP}\ }\textbf {\bibinfo
  {volume} {11}},\ \bibinfo {pages} {144}},\ \Eprint
  {https://arxiv.org/abs/1910.09567} {arXiv:1910.09567 [hep-ph]} \BibitemShut
  {NoStop}%
\bibitem [{\citenamefont {Arnold}\ \emph {et~al.}(2000)\citenamefont {Arnold},
  \citenamefont {Moore},\ and\ \citenamefont {Yaffe}}]{arnold_transport_2000}%
  \BibitemOpen
  \bibfield  {author} {\bibinfo {author} {\bibfnamefont {P.~B.}\ \bibnamefont
  {Arnold}}, \bibinfo {author} {\bibfnamefont {G.~D.}\ \bibnamefont {Moore}},\
  and\ \bibinfo {author} {\bibfnamefont {L.~G.}\ \bibnamefont {Yaffe}},\
  }\bibfield  {title} {\bibinfo {title} {{Transport coefficients in high
  temperature gauge theories. 1. Leading log results}},\ }\href
  {https://doi.org/10.1088/1126-6708/2000/11/001} {\bibfield  {journal}
  {\bibinfo  {journal} {JHEP}\ }\textbf {\bibinfo {volume} {11}},\ \bibinfo
  {pages} {001}},\ \Eprint {https://arxiv.org/abs/hep-ph/0010177}
  {arXiv:hep-ph/0010177} \BibitemShut {NoStop}%
\bibitem [{\citenamefont {Arnold}\ \emph {et~al.}(2003)\citenamefont {Arnold},
  \citenamefont {Moore},\ and\ \citenamefont {Yaffe}}]{arnold_transport_2003}%
  \BibitemOpen
  \bibfield  {author} {\bibinfo {author} {\bibfnamefont {P.~B.}\ \bibnamefont
  {Arnold}}, \bibinfo {author} {\bibfnamefont {G.~D.}\ \bibnamefont {Moore}},\
  and\ \bibinfo {author} {\bibfnamefont {L.~G.}\ \bibnamefont {Yaffe}},\
  }\bibfield  {title} {\bibinfo {title} {{Transport coefficients in high
  temperature gauge theories. 2. Beyond leading log}},\ }\href
  {https://doi.org/10.1088/1126-6708/2003/05/051} {\bibfield  {journal}
  {\bibinfo  {journal} {JHEP}\ }\textbf {\bibinfo {volume} {05}},\ \bibinfo
  {pages} {051}},\ \Eprint {https://arxiv.org/abs/hep-ph/0302165}
  {arXiv:hep-ph/0302165} \BibitemShut {NoStop}%
\bibitem [{\citenamefont {Greif}()}]{moritz_greif_electrical_2014}%
  \BibitemOpen
  \bibfield  {author} {\bibinfo {author} {\bibfnamefont {M.}~\bibnamefont
  {Greif}},\ }\emph {\bibinfo {title} {Electrical Conductivity of the Quark
  Gluon Plasma}},\ \href@noop {} {\bibinfo {type} {Master thesis}},\ \bibinfo
  {school} {Goethe University Frankfurt}\BibitemShut {NoStop}%
\bibitem [{\citenamefont {Aarts}\ and\ \citenamefont
  {Nikolaev}(2021)}]{aarts_electrical_2020}%
  \BibitemOpen
  \bibfield  {author} {\bibinfo {author} {\bibfnamefont {G.}~\bibnamefont
  {Aarts}}\ and\ \bibinfo {author} {\bibfnamefont {A.}~\bibnamefont
  {Nikolaev}},\ }\bibfield  {title} {\bibinfo {title} {{Electrical conductivity
  of the quark-gluon plasma: perspective from lattice QCD}},\ }\href
  {https://doi.org/10.1140/epja/s10050-021-00436-5} {\bibfield  {journal}
  {\bibinfo  {journal} {Eur. Phys. J. A}\ }\textbf {\bibinfo {volume} {57}},\
  \bibinfo {pages} {118} (\bibinfo {year} {2021})},\ \Eprint
  {https://arxiv.org/abs/2008.12326} {arXiv:2008.12326 [hep-lat]} \BibitemShut
  {NoStop}%
\bibitem [{\citenamefont {Brandt}\ \emph {et~al.}(2018)\citenamefont {Brandt},
  \citenamefont {Francis}, \citenamefont {Harris}, \citenamefont {Meyer},\ and\
  \citenamefont {Steinberg}}]{brandt_estimate_2018}%
  \BibitemOpen
  \bibfield  {author} {\bibinfo {author} {\bibfnamefont {B.~B.}\ \bibnamefont
  {Brandt}}, \bibinfo {author} {\bibfnamefont {A.}~\bibnamefont {Francis}},
  \bibinfo {author} {\bibfnamefont {T.}~\bibnamefont {Harris}}, \bibinfo
  {author} {\bibfnamefont {H.~B.}\ \bibnamefont {Meyer}},\ and\ \bibinfo
  {author} {\bibfnamefont {A.}~\bibnamefont {Steinberg}},\ }\bibfield  {title}
  {\bibinfo {title} {{An estimate for the thermal photon rate from lattice
  QCD}},\ }\href {https://doi.org/10.1051/epjconf/201817507044} {\bibfield
  {journal} {\bibinfo  {journal} {EPJ Web Conf.}\ }\textbf {\bibinfo {volume}
  {175}},\ \bibinfo {pages} {07044} (\bibinfo {year} {2018})},\ \Eprint
  {https://arxiv.org/abs/1710.07050} {arXiv:1710.07050 [hep-lat]} \BibitemShut
  {NoStop}%
\bibitem [{\citenamefont {Arnold}\ \emph {et~al.}(2001)\citenamefont {Arnold},
  \citenamefont {Moore},\ and\ \citenamefont {Yaffe}}]{Arnold:2001ms}%
  \BibitemOpen
  \bibfield  {author} {\bibinfo {author} {\bibfnamefont {P.~B.}\ \bibnamefont
  {Arnold}}, \bibinfo {author} {\bibfnamefont {G.~D.}\ \bibnamefont {Moore}},\
  and\ \bibinfo {author} {\bibfnamefont {L.~G.}\ \bibnamefont {Yaffe}},\
  }\bibfield  {title} {\bibinfo {title} {{Photon emission from quark gluon
  plasma: Complete leading order results}},\ }\href
  {https://doi.org/10.1088/1126-6708/2001/12/009} {\bibfield  {journal}
  {\bibinfo  {journal} {JHEP}\ }\textbf {\bibinfo {volume} {12}},\ \bibinfo
  {pages} {009}},\ \Eprint {https://arxiv.org/abs/hep-ph/0111107}
  {arXiv:hep-ph/0111107} \BibitemShut {NoStop}%
\bibitem [{\citenamefont {Aurenche}\ \emph {et~al.}(2002)\citenamefont
  {Aurenche}, \citenamefont {Gelis}, \citenamefont {Moore},\ and\ \citenamefont
  {Zaraket}}]{Aurenche:2002wq}%
  \BibitemOpen
  \bibfield  {author} {\bibinfo {author} {\bibfnamefont {P.}~\bibnamefont
  {Aurenche}}, \bibinfo {author} {\bibfnamefont {F.}~\bibnamefont {Gelis}},
  \bibinfo {author} {\bibfnamefont {G.~D.}\ \bibnamefont {Moore}},\ and\
  \bibinfo {author} {\bibfnamefont {H.}~\bibnamefont {Zaraket}},\ }\bibfield
  {title} {\bibinfo {title} {{Landau-Pomeranchuk-Migdal resummation for
  dilepton production}},\ }\href
  {https://doi.org/10.1088/1126-6708/2002/12/006} {\bibfield  {journal}
  {\bibinfo  {journal} {JHEP}\ }\textbf {\bibinfo {volume} {12}},\ \bibinfo
  {pages} {006}},\ \Eprint {https://arxiv.org/abs/hep-ph/0211036}
  {arXiv:hep-ph/0211036} \BibitemShut {NoStop}%
\bibitem [{\citenamefont {Carrington}\ \emph {et~al.}(2008)\citenamefont
  {Carrington}, \citenamefont {Gynther},\ and\ \citenamefont
  {Aurenche}}]{carrington_energetic_2008}%
  \BibitemOpen
  \bibfield  {author} {\bibinfo {author} {\bibfnamefont {M.~E.}\ \bibnamefont
  {Carrington}}, \bibinfo {author} {\bibfnamefont {A.}~\bibnamefont
  {Gynther}},\ and\ \bibinfo {author} {\bibfnamefont {P.}~\bibnamefont
  {Aurenche}},\ }\bibfield  {title} {\bibinfo {title} {{Energetic di-leptons
  from the Quark Gluon Plasma}},\ }\href
  {https://doi.org/10.1103/PhysRevD.77.045035} {\bibfield  {journal} {\bibinfo
  {journal} {Phys. Rev. D}\ }\textbf {\bibinfo {volume} {77}},\ \bibinfo
  {pages} {045035} (\bibinfo {year} {2008})},\ \Eprint
  {https://arxiv.org/abs/0711.3943} {arXiv:0711.3943 [hep-ph]} \BibitemShut
  {NoStop}%
\bibitem [{\citenamefont {Greif}\ \emph {et~al.}(2016)\citenamefont {Greif},
  \citenamefont {Greiner},\ and\ \citenamefont {Denicol}}]{Greif:2016skc}%
  \BibitemOpen
  \bibfield  {author} {\bibinfo {author} {\bibfnamefont {M.}~\bibnamefont
  {Greif}}, \bibinfo {author} {\bibfnamefont {C.}~\bibnamefont {Greiner}},\
  and\ \bibinfo {author} {\bibfnamefont {G.~S.}\ \bibnamefont {Denicol}},\
  }\bibfield  {title} {\bibinfo {title} {{Electric conductivity of a hot hadron
  gas from a kinetic approach}},\ }\href
  {https://doi.org/10.1103/PhysRevD.93.096012} {\bibfield  {journal} {\bibinfo
  {journal} {Phys. Rev. D}\ }\textbf {\bibinfo {volume} {93}},\ \bibinfo
  {pages} {096012} (\bibinfo {year} {2016})},\ \Eprint
  {https://arxiv.org/abs/1602.05085} {arXiv:1602.05085 [nucl-th]} \BibitemShut
  {NoStop}%
\bibitem [{\citenamefont {Yin}(2014)}]{yin_electrical_2014}%
  \BibitemOpen
  \bibfield  {author} {\bibinfo {author} {\bibfnamefont {Y.}~\bibnamefont
  {Yin}},\ }\bibfield  {title} {\bibinfo {title} {{Electrical conductivity of
  the quark-gluon plasma and soft photon spectrum in heavy-ion collisions}},\
  }\href {https://doi.org/10.1103/PhysRevC.90.044903} {\bibfield  {journal}
  {\bibinfo  {journal} {Phys. Rev. C}\ }\textbf {\bibinfo {volume} {90}},\
  \bibinfo {pages} {044903} (\bibinfo {year} {2014})},\ \Eprint
  {https://arxiv.org/abs/1312.4434} {arXiv:1312.4434 [nucl-th]} \BibitemShut
  {NoStop}%
\bibitem [{\citenamefont {Adare}\ \emph {et~al.}(2015)\citenamefont {Adare}
  \emph {et~al.}}]{PHENIX:2014nkk}%
  \BibitemOpen
  \bibfield  {author} {\bibinfo {author} {\bibfnamefont {A.}~\bibnamefont
  {Adare}} \emph {et~al.} (\bibinfo {collaboration} {PHENIX}),\ }\bibfield
  {title} {\bibinfo {title} {{Centrality dependence of low-momentum
  direct-photon production in Au$+$Au collisions at $\sqrt{s_{_{NN}}}=200$
  GeV}},\ }\href {https://doi.org/10.1103/PhysRevC.91.064904} {\bibfield
  {journal} {\bibinfo  {journal} {Phys. Rev. C}\ }\textbf {\bibinfo {volume}
  {91}},\ \bibinfo {pages} {064904} (\bibinfo {year} {2015})},\ \Eprint
  {https://arxiv.org/abs/1405.3940} {arXiv:1405.3940 [nucl-ex]} \BibitemShut
  {NoStop}%
\bibitem [{\citenamefont {Banerjee}\ \emph {et~al.}(2012)\citenamefont
  {Banerjee}, \citenamefont {Datta}, \citenamefont {Gavai},\ and\ \citenamefont
  {Majumdar}}]{banerjee_heavy_2012}%
  \BibitemOpen
  \bibfield  {author} {\bibinfo {author} {\bibfnamefont {D.}~\bibnamefont
  {Banerjee}}, \bibinfo {author} {\bibfnamefont {S.}~\bibnamefont {Datta}},
  \bibinfo {author} {\bibfnamefont {R.}~\bibnamefont {Gavai}},\ and\ \bibinfo
  {author} {\bibfnamefont {P.}~\bibnamefont {Majumdar}},\ }\bibfield  {title}
  {\bibinfo {title} {{Heavy Quark Momentum Diffusion Coefficient from Lattice
  QCD}},\ }\href {https://doi.org/10.1103/PhysRevD.85.014510} {\bibfield
  {journal} {\bibinfo  {journal} {Phys. Rev. D}\ }\textbf {\bibinfo {volume}
  {85}},\ \bibinfo {pages} {014510} (\bibinfo {year} {2012})},\ \Eprint
  {https://arxiv.org/abs/1109.5738} {arXiv:1109.5738 [hep-lat]} \BibitemShut
  {NoStop}%
\bibitem [{\citenamefont {Borsanyi}\ \emph {et~al.}(2012)\citenamefont
  {Borsanyi}, \citenamefont {Fodor}, \citenamefont {Katz}, \citenamefont
  {Krieg}, \citenamefont {Ratti},\ and\ \citenamefont
  {Szabo}}]{Borsanyi:2011sw}%
  \BibitemOpen
  \bibfield  {author} {\bibinfo {author} {\bibfnamefont {S.}~\bibnamefont
  {Borsanyi}}, \bibinfo {author} {\bibfnamefont {Z.}~\bibnamefont {Fodor}},
  \bibinfo {author} {\bibfnamefont {S.~D.}\ \bibnamefont {Katz}}, \bibinfo
  {author} {\bibfnamefont {S.}~\bibnamefont {Krieg}}, \bibinfo {author}
  {\bibfnamefont {C.}~\bibnamefont {Ratti}},\ and\ \bibinfo {author}
  {\bibfnamefont {K.}~\bibnamefont {Szabo}},\ }\bibfield  {title} {\bibinfo
  {title} {{Fluctuations of conserved charges at finite temperature from
  lattice QCD}},\ }\href {https://doi.org/10.1007/JHEP01(2012)138} {\bibfield
  {journal} {\bibinfo  {journal} {JHEP}\ }\textbf {\bibinfo {volume} {01}},\
  \bibinfo {pages} {138}},\ \Eprint {https://arxiv.org/abs/1112.4416}
  {arXiv:1112.4416 [hep-lat]} \BibitemShut {NoStop}%
\bibitem [{\citenamefont {Bazavov}\ \emph {et~al.}(2012)\citenamefont {Bazavov}
  \emph {et~al.}}]{bazavov_fluctuations_2012}%
  \BibitemOpen
  \bibfield  {author} {\bibinfo {author} {\bibfnamefont {A.}~\bibnamefont
  {Bazavov}} \emph {et~al.} (\bibinfo {collaboration} {HotQCD}),\ }\bibfield
  {title} {\bibinfo {title} {{Fluctuations and Correlations of net baryon
  number, electric charge, and strangeness: A comparison of lattice QCD results
  with the hadron resonance gas model}},\ }\href
  {https://doi.org/10.1103/PhysRevD.86.034509} {\bibfield  {journal} {\bibinfo
  {journal} {Phys. Rev. D}\ }\textbf {\bibinfo {volume} {86}},\ \bibinfo
  {pages} {034509} (\bibinfo {year} {2012})},\ \Eprint
  {https://arxiv.org/abs/1203.0784} {arXiv:1203.0784 [hep-lat]} \BibitemShut
  {NoStop}%
\bibitem [{\citenamefont {Nam}(2012)}]{nam_electrical_2012}%
  \BibitemOpen
  \bibfield  {author} {\bibinfo {author} {\bibfnamefont {S.-i.}\ \bibnamefont
  {Nam}},\ }\bibfield  {title} {\bibinfo {title} {{Electrical conductivity of
  quark matter at finite T under external magnetic field}},\ }\href
  {https://doi.org/10.1103/PhysRevD.86.033014} {\bibfield  {journal} {\bibinfo
  {journal} {Phys. Rev. D}\ }\textbf {\bibinfo {volume} {86}},\ \bibinfo
  {pages} {033014} (\bibinfo {year} {2012})},\ \Eprint
  {https://arxiv.org/abs/1207.3172} {arXiv:1207.3172 [hep-ph]} \BibitemShut
  {NoStop}%
\bibitem [{\citenamefont {Hattori}\ and\ \citenamefont
  {Satow}(2016)}]{hattori_electrical_2016}%
  \BibitemOpen
  \bibfield  {author} {\bibinfo {author} {\bibfnamefont {K.}~\bibnamefont
  {Hattori}}\ and\ \bibinfo {author} {\bibfnamefont {D.}~\bibnamefont
  {Satow}},\ }\bibfield  {title} {\bibinfo {title} {{Electrical Conductivity of
  Quark-Gluon Plasma in Strong Magnetic Fields}},\ }\href
  {https://doi.org/10.1103/PhysRevD.94.114032} {\bibfield  {journal} {\bibinfo
  {journal} {Phys. Rev. D}\ }\textbf {\bibinfo {volume} {94}},\ \bibinfo
  {pages} {114032} (\bibinfo {year} {2016})},\ \Eprint
  {https://arxiv.org/abs/1610.06818} {arXiv:1610.06818 [hep-ph]} \BibitemShut
  {NoStop}%
\bibitem [{\citenamefont {Feng}(2017)}]{feng_electric_2017}%
  \BibitemOpen
  \bibfield  {author} {\bibinfo {author} {\bibfnamefont {B.}~\bibnamefont
  {Feng}},\ }\bibfield  {title} {\bibinfo {title} {{Electric conductivity and
  Hall conductivity of the QGP in a magnetic field}},\ }\href
  {https://doi.org/10.1103/PhysRevD.96.036009} {\bibfield  {journal} {\bibinfo
  {journal} {Phys. Rev. D}\ }\textbf {\bibinfo {volume} {96}},\ \bibinfo
  {pages} {036009} (\bibinfo {year} {2017})}\BibitemShut {NoStop}%
\bibitem [{\citenamefont {Floerchinger}\ \emph {et~al.}(2019)\citenamefont
  {Floerchinger}, \citenamefont {Grossi},\ and\ \citenamefont
  {Lion}}]{floerchinger_fluid_2019}%
  \BibitemOpen
  \bibfield  {author} {\bibinfo {author} {\bibfnamefont {S.}~\bibnamefont
  {Floerchinger}}, \bibinfo {author} {\bibfnamefont {E.}~\bibnamefont
  {Grossi}},\ and\ \bibinfo {author} {\bibfnamefont {J.}~\bibnamefont {Lion}},\
  }\bibfield  {title} {\bibinfo {title} {{Fluid dynamics of heavy ion
  collisions with mode expansion}},\ }\href
  {https://doi.org/10.1103/PhysRevC.100.014905} {\bibfield  {journal} {\bibinfo
   {journal} {Phys. Rev. C}\ }\textbf {\bibinfo {volume} {100}},\ \bibinfo
  {pages} {014905} (\bibinfo {year} {2019})},\ \Eprint
  {https://arxiv.org/abs/1811.01870} {arXiv:1811.01870 [nucl-th]} \BibitemShut
  {NoStop}%
\bibitem [{\citenamefont {Devetak}\ \emph {et~al.}(2020)\citenamefont
  {Devetak}, \citenamefont {Dubla}, \citenamefont {Floerchinger}, \citenamefont
  {Grossi}, \citenamefont {Masciocchi}, \citenamefont {Mazeliauskas},\ and\
  \citenamefont {Selyuzhenkov}}]{Devetak:2019lsk}%
  \BibitemOpen
  \bibfield  {author} {\bibinfo {author} {\bibfnamefont {D.}~\bibnamefont
  {Devetak}}, \bibinfo {author} {\bibfnamefont {A.}~\bibnamefont {Dubla}},
  \bibinfo {author} {\bibfnamefont {S.}~\bibnamefont {Floerchinger}}, \bibinfo
  {author} {\bibfnamefont {E.}~\bibnamefont {Grossi}}, \bibinfo {author}
  {\bibfnamefont {S.}~\bibnamefont {Masciocchi}}, \bibinfo {author}
  {\bibfnamefont {A.}~\bibnamefont {Mazeliauskas}},\ and\ \bibinfo {author}
  {\bibfnamefont {I.}~\bibnamefont {Selyuzhenkov}},\ }\bibfield  {title}
  {\bibinfo {title} {{Global fluid fits to identified particle transverse
  momentum spectra from heavy-ion collisions at the Large Hadron Collider}},\
  }\href {https://doi.org/10.1007/JHEP06(2020)044} {\bibfield  {journal}
  {\bibinfo  {journal} {JHEP}\ }\textbf {\bibinfo {volume} {06}},\ \bibinfo
  {pages} {044}},\ \Eprint {https://arxiv.org/abs/1909.10485} {arXiv:1909.10485
  [hep-ph]} \BibitemShut {NoStop}%
\bibitem [{\citenamefont {Bernhard}\ \emph {et~al.}(2016)\citenamefont
  {Bernhard}, \citenamefont {Moreland}, \citenamefont {Bass}, \citenamefont
  {Liu},\ and\ \citenamefont {Heinz}}]{Bernhard:2016tnd}%
  \BibitemOpen
  \bibfield  {author} {\bibinfo {author} {\bibfnamefont {J.~E.}\ \bibnamefont
  {Bernhard}}, \bibinfo {author} {\bibfnamefont {J.~S.}\ \bibnamefont
  {Moreland}}, \bibinfo {author} {\bibfnamefont {S.~A.}\ \bibnamefont {Bass}},
  \bibinfo {author} {\bibfnamefont {J.}~\bibnamefont {Liu}},\ and\ \bibinfo
  {author} {\bibfnamefont {U.}~\bibnamefont {Heinz}},\ }\bibfield  {title}
  {\bibinfo {title} {{Applying Bayesian parameter estimation to relativistic
  heavy-ion collisions: simultaneous characterization of the initial state and
  quark-gluon plasma medium}},\ }\href
  {https://doi.org/10.1103/PhysRevC.94.024907} {\bibfield  {journal} {\bibinfo
  {journal} {Phys. Rev. C}\ }\textbf {\bibinfo {volume} {94}},\ \bibinfo
  {pages} {024907} (\bibinfo {year} {2016})},\ \Eprint
  {https://arxiv.org/abs/1605.03954} {arXiv:1605.03954 [nucl-th]} \BibitemShut
  {NoStop}%
\bibitem [{\citenamefont {Turbide}\ \emph {et~al.}(2004)\citenamefont
  {Turbide}, \citenamefont {Rapp},\ and\ \citenamefont
  {Gale}}]{Turbide:2003si}%
  \BibitemOpen
  \bibfield  {author} {\bibinfo {author} {\bibfnamefont {S.}~\bibnamefont
  {Turbide}}, \bibinfo {author} {\bibfnamefont {R.}~\bibnamefont {Rapp}},\ and\
  \bibinfo {author} {\bibfnamefont {C.}~\bibnamefont {Gale}},\ }\bibfield
  {title} {\bibinfo {title} {{Hadronic production of thermal photons}},\ }\href
  {https://doi.org/10.1103/PhysRevC.69.014903} {\bibfield  {journal} {\bibinfo
  {journal} {Phys. Rev. C}\ }\textbf {\bibinfo {volume} {69}},\ \bibinfo
  {pages} {014903} (\bibinfo {year} {2004})},\ \Eprint
  {https://arxiv.org/abs/hep-ph/0308085} {arXiv:hep-ph/0308085} \BibitemShut
  {NoStop}%
\bibitem [{\citenamefont {Mazeliauskas}\ \emph {et~al.}(2019)\citenamefont
  {Mazeliauskas}, \citenamefont {Floerchinger}, \citenamefont {Grossi},\ and\
  \citenamefont {Teaney}}]{mazeliauskas_fast_2019}%
  \BibitemOpen
  \bibfield  {author} {\bibinfo {author} {\bibfnamefont {A.}~\bibnamefont
  {Mazeliauskas}}, \bibinfo {author} {\bibfnamefont {S.}~\bibnamefont
  {Floerchinger}}, \bibinfo {author} {\bibfnamefont {E.}~\bibnamefont
  {Grossi}},\ and\ \bibinfo {author} {\bibfnamefont {D.}~\bibnamefont
  {Teaney}},\ }\bibfield  {title} {\bibinfo {title} {{Fast resonance decays in
  nuclear collisions}},\ }\href
  {https://doi.org/10.1140/epjc/s10052-019-6791-7} {\bibfield  {journal}
  {\bibinfo  {journal} {Eur. Phys. J. C}\ }\textbf {\bibinfo {volume} {79}},\
  \bibinfo {pages} {284} (\bibinfo {year} {2019})},\ \Eprint
  {https://arxiv.org/abs/1809.11049} {arXiv:1809.11049 [nucl-th]} \BibitemShut
  {NoStop}%
\bibitem [{\citenamefont {Sjostrand}\ \emph {et~al.}(2008)\citenamefont
  {Sjostrand}, \citenamefont {Mrenna},\ and\ \citenamefont
  {Skands}}]{Sjostrand:2007gs}%
  \BibitemOpen
  \bibfield  {author} {\bibinfo {author} {\bibfnamefont {T.}~\bibnamefont
  {Sjostrand}}, \bibinfo {author} {\bibfnamefont {S.}~\bibnamefont {Mrenna}},\
  and\ \bibinfo {author} {\bibfnamefont {P.~Z.}\ \bibnamefont {Skands}},\
  }\bibfield  {title} {\bibinfo {title} {{A Brief Introduction to PYTHIA
  8.1}},\ }\href {https://doi.org/10.1016/j.cpc.2008.01.036} {\bibfield
  {journal} {\bibinfo  {journal} {Comput. Phys. Commun.}\ }\textbf {\bibinfo
  {volume} {178}},\ \bibinfo {pages} {852} (\bibinfo {year} {2008})},\ \Eprint
  {https://arxiv.org/abs/0710.3820} {arXiv:0710.3820 [hep-ph]} \BibitemShut
  {NoStop}%
\bibitem [{\citenamefont {Albrecht}\ \emph {et~al.}(1995)\citenamefont
  {Albrecht} \emph {et~al.}}]{WA80:1995whm}%
  \BibitemOpen
  \bibfield  {author} {\bibinfo {author} {\bibfnamefont {R.}~\bibnamefont
  {Albrecht}} \emph {et~al.} (\bibinfo {collaboration} {WA80}),\ }\bibfield
  {title} {\bibinfo {title} {{Production of eta mesons in 200-A/GeV S + S and S
  + Au reactions}},\ }\href {https://doi.org/10.1016/0370-2693(95)01166-N}
  {\bibfield  {journal} {\bibinfo  {journal} {Phys. Lett. B}\ }\textbf
  {\bibinfo {volume} {361}},\ \bibinfo {pages} {14} (\bibinfo {year} {1995})},\
  \Eprint {https://arxiv.org/abs/hep-ex/9507009} {arXiv:hep-ex/9507009}
  \BibitemShut {NoStop}%
\bibitem [{\citenamefont {Kroll}\ and\ \citenamefont
  {Wada}(1955)}]{Kroll:1955zu}%
  \BibitemOpen
  \bibfield  {author} {\bibinfo {author} {\bibfnamefont {N.~M.}\ \bibnamefont
  {Kroll}}\ and\ \bibinfo {author} {\bibfnamefont {W.}~\bibnamefont {Wada}},\
  }\bibfield  {title} {\bibinfo {title} {{Internal pair production associated
  with the emission of high-energy gamma rays}},\ }\href
  {https://doi.org/10.1103/PhysRev.98.1355} {\bibfield  {journal} {\bibinfo
  {journal} {Phys. Rev.}\ }\textbf {\bibinfo {volume} {98}},\ \bibinfo {pages}
  {1355} (\bibinfo {year} {1955})}\BibitemShut {NoStop}%
\bibitem [{\citenamefont {Landsberg}(1985)}]{Landsberg:1985gaz}%
  \BibitemOpen
  \bibfield  {author} {\bibinfo {author} {\bibfnamefont {L.~G.}\ \bibnamefont
  {Landsberg}},\ }\bibfield  {title} {\bibinfo {title} {{Electromagnetic Decays
  of Light Mesons}},\ }\href {https://doi.org/10.1016/0370-1573(85)90129-2}
  {\bibfield  {journal} {\bibinfo  {journal} {Phys. Rept.}\ }\textbf {\bibinfo
  {volume} {128}},\ \bibinfo {pages} {301} (\bibinfo {year}
  {1985})}\BibitemShut {NoStop}%
\bibitem [{\citenamefont {Aggarwal}\ \emph {et~al.}(2004)\citenamefont
  {Aggarwal} \emph {et~al.}}]{WA98:2003ukc}%
  \BibitemOpen
  \bibfield  {author} {\bibinfo {author} {\bibfnamefont {M.~M.}\ \bibnamefont
  {Aggarwal}} \emph {et~al.} (\bibinfo {collaboration} {WA98}),\ }\bibfield
  {title} {\bibinfo {title} {{Interferometry of direct photons in central
  Pb-208+Pb-208 collisions at 158-A-GeV}},\ }\href
  {https://doi.org/10.1103/PhysRevLett.93.022301} {\bibfield  {journal}
  {\bibinfo  {journal} {Phys. Rev. Lett.}\ }\textbf {\bibinfo {volume} {93}},\
  \bibinfo {pages} {022301} (\bibinfo {year} {2004})},\ \Eprint
  {https://arxiv.org/abs/nucl-ex/0310022} {arXiv:nucl-ex/0310022} \BibitemShut
  {NoStop}%
\bibitem [{\citenamefont {Peressounko}(2003)}]{Peressounko:2003cf}%
  \BibitemOpen
  \bibfield  {author} {\bibinfo {author} {\bibfnamefont {D.}~\bibnamefont
  {Peressounko}},\ }\bibfield  {title} {\bibinfo {title} {{Hanbury Brown-Twiss
  interferometry of direct photons in heavy ion collisions}},\ }\href
  {https://doi.org/10.1103/PhysRevC.67.014905} {\bibfield  {journal} {\bibinfo
  {journal} {Phys. Rev. C}\ }\textbf {\bibinfo {volume} {67}},\ \bibinfo
  {pages} {014905} (\bibinfo {year} {2003})}\BibitemShut {NoStop}%
\bibitem [{\citenamefont {Fernandez-Fraile}\ and\ \citenamefont
  {Gomez~Nicola}(2009)}]{Fernandez-Fraile:2009eug}%
  \BibitemOpen
  \bibfield  {author} {\bibinfo {author} {\bibfnamefont {D.}~\bibnamefont
  {Fernandez-Fraile}}\ and\ \bibinfo {author} {\bibfnamefont {A.}~\bibnamefont
  {Gomez~Nicola}},\ }\bibfield  {title} {\bibinfo {title} {{Transport
  coefficients and resonances for a meson gas in Chiral Perturbation Theory}},\
  }\href {https://doi.org/10.1140/epjc/s10052-009-0935-0} {\bibfield  {journal}
  {\bibinfo  {journal} {Eur. Phys. J. C}\ }\textbf {\bibinfo {volume} {62}},\
  \bibinfo {pages} {37} (\bibinfo {year} {2009})},\ \Eprint
  {https://arxiv.org/abs/0902.4829} {arXiv:0902.4829 [hep-ph]} \BibitemShut
  {NoStop}%
\bibitem [{\citenamefont {Adamov\'a}\ \emph {et~al.}(2019)\citenamefont
  {Adamov\'a} \emph {et~al.}}]{Adamova:2019vkf}%
  \BibitemOpen
  \bibfield  {author} {\bibinfo {author} {\bibfnamefont {D.}~\bibnamefont
  {Adamov\'a}} \emph {et~al.},\ }\bibfield  {title} {\bibinfo {title} {{A
  next-generation LHC heavy-ion experiment}},\ }\href@noop {} {\  (\bibinfo
  {year} {2019})},\ \Eprint {https://arxiv.org/abs/1902.01211}
  {arXiv:1902.01211 [physics.ins-det]} \BibitemShut {NoStop}%
\bibitem [{\citenamefont {Floerchinger}(2016)}]{floerchinger_variational_2016}%
  \BibitemOpen
  \bibfield  {author} {\bibinfo {author} {\bibfnamefont {S.}~\bibnamefont
  {Floerchinger}},\ }\bibfield  {title} {\bibinfo {title} {{Variational
  principle for theories with dissipation from analytic continuation}},\ }\href
  {https://doi.org/10.1007/JHEP09(2016)099} {\bibfield  {journal} {\bibinfo
  {journal} {JHEP}\ }\textbf {\bibinfo {volume} {09}},\ \bibinfo {pages}
  {099}},\ \Eprint {https://arxiv.org/abs/1603.07148} {arXiv:1603.07148
  [hep-th]} \BibitemShut {NoStop}%
\end{thebibliography}%
\end{document}